\documentclass[11pt]{article}
\usepackage{times}
\usepackage{cospar}
\usepackage{graphicx}
\usepackage[sectionbib]{natbib}

\title{THE STATE OF PULSAR THEORY}

\author{F. C. Michel\address{Physics and Astronomy, Rice University, Houston, TX 77005, USA}}

\begin{document}

\maketitle

\begin{abstract}
I summarize the status of pulsar theory, now 35 years after their
discovery.  Although progress has been made in understanding the
relevant processes involved, there are several widely held
misconceptions that are inhibiting further advances.  These include
the idea that plasma ``must'' be accelerated from the magnetic polar
caps (the basis for the ``Hollow Cone Model'') and the idea that winds
would be driven away by centrifugal forces, with large amplitude
electromagnetic waves playing no role whatsoever.  However, recent
theoretical work is converging on a picture that closely resembles the
latest HST and CHANDRA images, providing hope for the future.  No less
than 3 groups have recently confirmed the early 
Krause-Polstorff-Michel simulations
showing that the fundamental plasma distribution around a rotating
neutron star consists of two polar domes and an equatorial torus of
trapped nonneutral plasma of opposite sign charges.  Unless a lot of
new physics can be added, this distribution renders the
Goldreich-Julian
model irrelevant (i.e., along with most of the theoretical publications over 
the last 33 years).

\end{abstract}

\section{HISTORICAL NOTES}
\vspace{2mm}
Pulsar research is probably similar to a number of other fields,
but it definitely has some distinctive characteristics.
One characteristic, representative of the theorists, was
shaped by the shockingly unexpected discovery of the radio
pulsar phenomenon.  Compact sources of intense radio emission
were about the last thing theorists expected to be discovered.
This led to the general expectation that the mechanism would
have to be so extraordinary that, once exposed, would be
obvious for all to see.  This in turn lead to a feverish ``race''
to explain pulsars, with Tommy Gold being in the forefront with
his model of bunches of charged particles being swung about a
magnetized neutron star, and being confined out to the ``light
circle.''  I personally was bombarded almost weekly with
revised versions of another theory from a pair of theorists.
This highly competitive atmosphere produced a rather disruptive
research atmosphere where everyone went their own direction,
which I will call here, for lack of a very clever name,

{\bf1. Searching for the quick kill (theorists).}
This general atmosphere probably resulted from people being drawn from
diverse research backgrounds, who had therefore never collaborated
before.  Nevertheless, in fairly quick order the likely nature of the
astrophysical object involved narrowed down to the rotating,
magnetized, neutron star model of today.  The major theoretical
dispute was over the possibility that the stellar object might be a
known object (a white dwarf, as favored by the ``smart money'' of the
time) rather than the new and therefore exotic object, the neutron
star.

Observers, on the other hand, while undoubtedly scrambling to be
the first to discover the next pulsar (especially when the list
stalled at four for a while), nevertheless seemed to be rather
more collegial.  And the observers
seemed to believe that, with sufficient observation and analysis,
they would be able to figure out how pulsars worked from observation
alone.  One might call this,

{\bf 2. Maintaining community solidarity (observers).}  After all, the
radio astronomy community had been around for some time prior to the
discovery of pulsars, and ways of working together had been developed.
The {\em relatively} cooperative and congenial attitude among observers can
be illustrated by the idea that the magnetic fields of pulsars decayed
with a period of a few million years.  This idea was suggested around
1975 and pretty well became the accepted view among observers until
perhaps 1995, even though few observers actually checked over the data
to see if this claim could be supported.  Finally, straightforward
simulations
showed that the data did not support magnetic field decay on that
short a time scale (see e.g. Michel, 1991a).  It is curious
that among the physics community, virtually no one believed in this
rapid magnetic field decay, although there are to this day those in
the astrophysical community that adore the idea (since it is, after
all, plausible).  The fact that this idea was so widely accepted in
the observational community for so long illustrates the supportive
atmosphere at work.

Another set of characteristics seem relevant.  Among the theorists,
a corollary to the first attitude is to

{\bf 3. Exaggerate even the smallest incremental advance.}
It should be clear that this tendency is a direct consequence of the
idea that the theoretical resolution of how pulsars work would be
right around the corner.  So ``rounding'' any corner naturally sparked
huge excitement.  And every such small advance usually lead to a
publication announcing (``at last'') the definitive understanding of
pulsars.

Meanwhile, the observational community came to be

{\bf 4. Indifferent to theoretical disputes.}
Given the disarray among theorists, with each following their own
idiosyncratic ideas, the observational community has understandably
grown to think that little profit is to be had from sorting through
the various disagreements among theorists.  And this too is consistent
with their underlying assumption that they would succeed in understanding
how pulsars work on the basis of observation alone.

If one made a parallel to pulsar research with other research efforts
like, say, the ``war on cancer,'' which seems to have dragged on a
similar length of time, it would seem bizarre.  Here we would have a
handful of people (``theorists'') who sit in their offices with
pencil and paper trying to figure out how cancer works, only one or
two of whom would be funded at any one time, compared with numerous
doctors (``observers'') who insist that, if they only could
see enough patients, they would be able to figure out how cancer
worked by themselves.

Finally, we should note that despite their importance, 

{\bf 5. Much of pulsar data are private.}
This contrasts with NASA sponsored research where spacecraft data is
usually made available within a few years (ideally one year) of
acquisition.  But NASA does not run radio telescopes.  I am not sure
what the current situation is, but from experience, it has not been
easy (i.e., possible) for most theorists to readily obtain and analyze
radio observations of pulsars (I'll now have to start to say ``radio
pulsars'').

In this regard I cannot help but think of E. K. Bigg, whose name
will not probably ring a bell for many readers, but what he
noticed may ring a bell.  In 1964, Bigg published the astonishing
observation that Jupiter's decametric radiation was modulated by its
innermost satellite Io.  He discovered this entirely by fiddling with the
data.  It was entirely unexpected theoretically and seemed implausible to
the observers.

But there will not likely be such a breakthrough for pulsars, since
pulsar data is not only not available to amateurs, it is largely
unavailable to theorists.  Not that theorists necessarily have any
special expertise.  It is quite the opposite.  Sometimes someone needs
to ask a stupid question (could Io possibly effect radio emission from
an entire planet, with the largest magnetosphere in the entire solar
system?).  The relative lack of interest by observers in theory works to a
similar disadvantage.  Smart people are not looking over these
shoulders.  No one will say, ``you forgot to carry the one.''  Of course
if theory were to predict some definite correlation or the like,
certainly observers would check it.  But they are unlikely to go out
of their way to check the theoretical underpinning.  In other fields the
experimentalists are usually as well informed as the theorists.  It is
just a division of labor.

I conclude this section with the following observation:
$$
1967 + 35 = 2002.
$$
So 35 years have passed.  Will the way pulsars work yield to theory
alone?  Will it yield to observation alone?  The numbers are not on
anyone's side.  I will try to suggest why theory hasn't done it yet.
One reason was the chaotic approach to the problem, everyone running
in different directions.  The other is, paradoxically, how a strong
line of conventional wisdom keeps people from looking in
different directions, which we will next examine.  

\section{PULSAR RESEARCH IS NO LONGER ``NEW''}
\vspace{2mm}
It is worth noting that 35 years significantly exceeds the period
where a topic can be safely considered ``new.''  One attraction of a
brand new field is that one does not have to do much literature
search.  Almost any idea will be a new idea in the sense that no one
was thinking about this topic before.  However, after 35 years, the
chances are high that any ``new'' idea will not actually be new.

{\bf 6. No one can correctly assume that their ``new'' idea is
actually new.}  In fact, this whole attitude was not strictly true
even on ``day one,'' since related research areas may already have
inspired related theoretical work.  Examples relevant to pulsars are
Deutsch (1955), who published the electromagnetic fields expected
surrounding a rotating magnetized star, and Piddington (1957), who
suggested that the magnetic field of the nebula might be that of the
wound-up field of a rotating central star.  Of particular note is the
paper by Hones and Bergeson (1965) who show that a rotating magnetized
body will be surrounded by a plasma density

\begin{equation}
\rho = - \frac{\mu \omega}{2 \pi c r^3} 
\left\{
\left[ 3 \cos^2 \theta - 1 \right] \cos\gamma
+ 
\left[ 3 \sin \theta \cos \theta \cos ( \phi - \omega t) \right] \sin \gamma .
\right\}
\end{equation}
(their equation 17), where $\gamma$ is the inclination angle of the
dipole magnetic field.  Here $\mu$ is the magnetic moment, $\omega$ is
the stellar rotation rate, and the angles are the usual spherical
angles.  Their expression is therefore completely general and not
confined to the special case that the rotation vector is aligned with
the magnetization axis.  And they point out that the consequence is
that this plasma will rigidly rotate with the body (star, say).
Indeed, they point out that these deductions (rigid corotation and a
specific space charge density) were made much earlier by L. Davis,
Jr. (1947, 1948).  Thus the ``modern'' view that rotating magnetized
neutron stars would be surrounded by plasma, for example, is anything
but new.  Four years after Hones and Bergeson, Goldreich and Julian
(1969) proposed the same thing in regard to pulsars, restricting
themselves to the trivial case of an aligned rotator, and adding the
assumption that a wind would be generated at the light cylinder owing
to centrifugal forces due to the rigid corotation.  Unfortunately,
this one innovation proves difficult to verify.  Equally unverified is
the supposition that such a wind demands in turn the replacement by
new particles accelerated out of the magnetic polar caps.

\subsection{Consequences}
Nevertheless, these ideas permeate the pulsar field, possibly because
the magnetic polar caps are clearly special places and therefore
plausible sites of special activity.  But the underlying model offered
rather little in the way of explaining how pulsars might work.

$\bullet$ No pulses were actually produced (the model being restricted to the aligned
case, although Hones and Bergeson gave solutions for the general case of
arbitrary inclination).

$\bullet$ No reason for radio emission was supplied.

$\bullet$ No reason for coherent emission was supplied.

$\bullet$ No idea of the spectral distribution was supplied.

$\bullet$ No idea of any high-energy emissions was supplied.

$\bullet$ The model was unable to provide closed electric currents.

$\bullet$ The model was in fact shown to be incorrect (1980-1985)

The model was nevertheless quite popular since it was

$\bullet$ Easy to understand, and

$\bullet$ ``Self-consistent.''

\subsection{Polar Cap Acceleration (Observation)}

The commonly accepted idea is that radio emission is beamed outwards
from the magnetic polar caps, that must then be inclined relative to
the spin axis of the neutron star so as to sweep a ``lighthouse''
beam through the sky and produce brief pulses as seen by distant observers.

\begin{figure}[t]
\begin{minipage}{90mm}
\includegraphics[width=75mm]{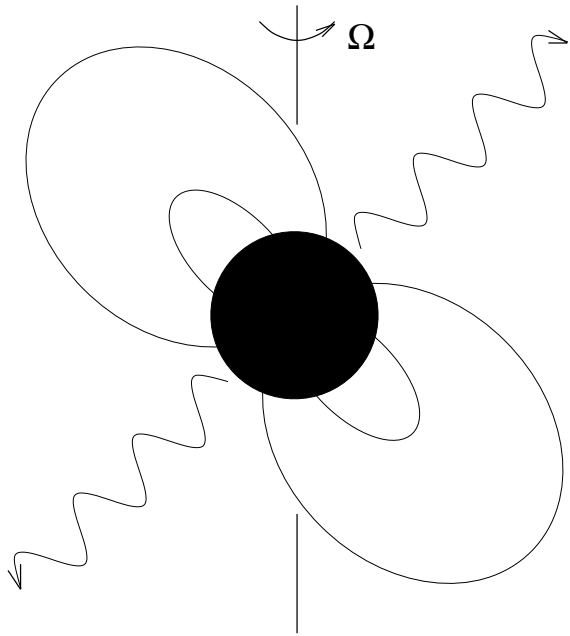}
\end{minipage}
\hfil\hspace{\fill}
\begin{minipage}{75mm}
\caption{Tilting the aligned
rotator to get pulses.  Wiggly lines
represent the supposed origin of radio emission from the magnetic polar
caps.  Thin lines represent closed magnetic field lines near the neutron
star (black sphere).  Although clearly a generic model, the assumption here
that the radiation is axially aligned with a dipole moment is manifest.}
\label{wigfig}
\end{minipage}
\end{figure}

This is the so-called ``Hollow Cone Model,'' based on this idea that
particles are being accelerated out of the polar caps to replace those
lost by a centrifugal wind at the light cylinder.  This model was and
is a very popular one among observers, and in fact it is used implicitly
to infer the angle between the spin and magnetization axis as well as
that between the spin and line-of-sight to the observer (usually
denoted ``$\alpha$'' and ``$\beta$'').  These angles are inferred by
assuming that the magnetic field lines radiate radially (as projected
on the surface) and that the polarization angle projected on the sky
is fixed relative to the direction of those lines.  Usually it is
assumed that the polarization is either parallel or perpendicular, but
orthogonal polarization jumps complicate the interpretation some.
Figure~1 is now in many popular books, so it will probably be impossible
to get this corrected even if, as we strongly suspect, this whole story
turns out to be wrong.

\subsection{Polar Cap Acceleration (Theory)}

Exactly the same assumption has entered the theoretical work.
If putative replacement from the magnetic polar caps is where the
radiation comes from, then one should be able to zoom in on just the
polar caps and ignore the rest of the system, Figure 2.
Many theorists thereby ignored
any aspect of the physics except that which might take place over the
polar caps.  But explicit simulation shows that there are no forces
excited by this loss of plasma.

\begin{figure}[b]
\begin{minipage}{90mm}
\includegraphics[width=75mm]{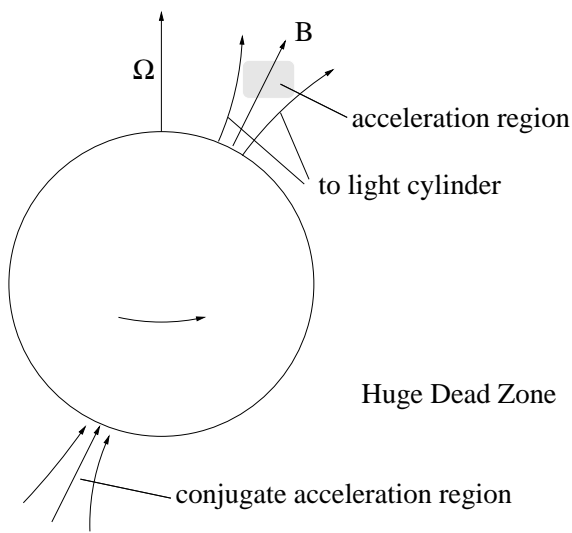}
\end{minipage}
\hfil\hspace{\fill}
\begin{minipage}{75mm}
\caption{\hspace{.1in}Generic polar cap
 acceleration region, adapted from
Michel (1975).  The locus of field lines
to the light-cylinder is taken to demark
the polar cap which supposedly encompasses
the field lines on which particles from the surface
are accelerated.  Remaining closed field lines
enclose a huge corotating inactive zone.}
\label{fig2}
\end{minipage}
\end{figure}

\subsection{``Light Cylinder''}

Among many curiosities is the endurance of the term, ``Light
Cylinder,'' which was actually coined by T. Gold (1969) in connection
with a model that no one believes any more.  In that model, Gold had
an ad hoc bunch of charged particles swung in a circle as fast as
possible, namely at a distance $r = c/\Omega$ from the rotating star
(we use an upper case $\Omega$ to underscore that this is a
macroscopic quantity, unlike Hones and Bergeson (HB), who use $\omega$.  
Goldreich and
Julian (GJ) assumed that a plasma wind transition took place at the same
distance and this term persisted, although it carries with it the
direct implication that the physics of the pulsar system is somehow
dominated by centrifugal effects above all else.  In the same way, it
has constantly been assumed in phenomenological papers that the
magnetic field lines that went ``to'' the light cylinder were of
special character.  Typically the ``Hollow Cone'' in the model
pertains to the locus of those magnetic field lines that reach out to
the light cylinder.  From symmetry, this would define a circle (the
magnetic polar cap) for aligned rotation.  Some papers try to
``correct'' for the shape of the polar caps when the magnetic dipole
is inclined, since now a different set of field lines would be
involved depending on the azimuthal direction of the field line, as
shown in Figure~3.

\begin{figure}[t]
\begin{minipage}{90mm}
\includegraphics[width=75mm]{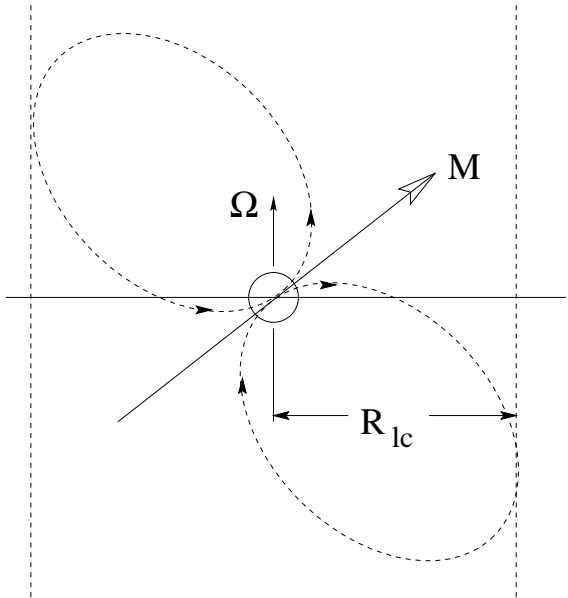}
\end{minipage}
\hfil\hspace{\fill}
\begin{minipage}{75mm}
\caption{Magical magnetic
field lines that touch the light cylinder, adapted from
Michel (1991a).  Note the disparity in lengths of the field
lines to the opposite sides of the polar caps.} 
\label{fig:litecyl}
\end{minipage}
\end{figure}

\section{The Polar Cap Acceleration Myth}

What we have been discussing is the polar cap acceleration ``myth.''
It is a myth not because we can disprove that the
radiation comes from the polar caps.  It is a myth because
the model that generated this idea is baseless.

\begin{figure}[b]
\begin{minipage}{90mm}
\includegraphics[width=75mm]{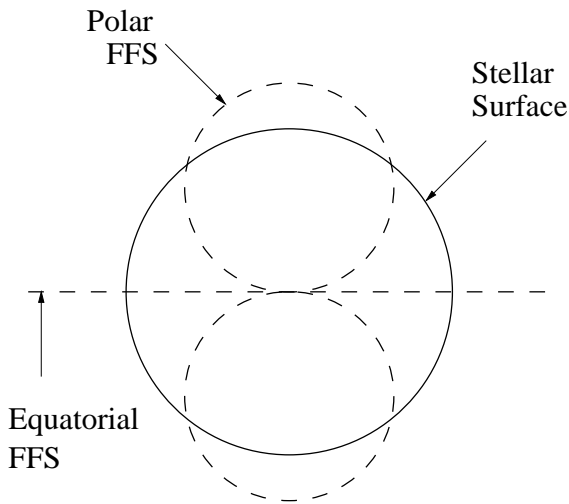}
\end{minipage}
\hfil\hspace{\fill}
\begin{minipage}{75mm}
\caption{\hspace{.1in}Trapping surfaces of an
aligned rotator, denoted
here ``FFS'' for Force-Free Surface, from Michel and Pellat
(1981).  The {\bf E} field component parallel to {\bf B}
reverses sign at an FFS, and therefore particles of one
sign are pushed together (trapped) here, while the other sign
are pushed away.
The surfaces outside are independent of the surfaces
inside, but are continuous as shown if
the magnetic field is from a point dipole at
the center.} 
\label{fig:ffs}
\end{minipage}
\end{figure}

Figure 4 shows the electrostatics above the polar cap of an
aligned rotator.  The basic electrostatics is simple to understand.
First, it is quite true that an aligned rotator will act to pull charged
particles up off the polar caps (and much of the rest of the star).
But this is not the end of the story.  It is easy to show that an aligned
rotator will have an intrinsic electrostatic charge.  For simplicity this
charge is shown concentrated at the center of the star, although the exact
distribution depends on the actual magnetic field structure inside the
star.  For a point dipole all the way to the center, the charge will be all
at the center, which is the simplest case.  The more complicated cases
simply obscure the essential details that follow.  The basic electrostatics
is explained in detail in Michel (1991a) and in even more detail in
Michel and Li (1999).

What the central charge does is attract back toward the star exactly
the sign of charge that would be emitted over the polar caps.
Basically, the field that leads to emission is quadrupolar in nature.
This field is highly effective close to the surface, but falls off
rapidly ($1/r^4$) and is overtaken by the monopole field ($1/r^2$) of the
central charge.  Thus particles moving along magnetic field lines {\em
are} accelerated off the surface but they pass through a trapping
surface and above that surface they are accelerated back towards the polar
caps.  By failing to notice this fundamental feature of the
electrostatics of the aligned rotator, enthusiasts have assumed that
the acceleration would act to infinity.  In Figure~4 , the locus
where the two balance is labeled
``FFS Polar,'' is drawn to scale, and shows how small the actual acceleration
region is.  The trapped particles accumulate on this force-free surface
(FFS).

The contrary is easy is illustrate.  In Figure~5 we sketch how an
aligned rotator would be charged if it were a rotating sphere of laboratory
dimensions.

\begin{figure}[h]
\begin{minipage}{90mm}
\includegraphics[width=75mm]{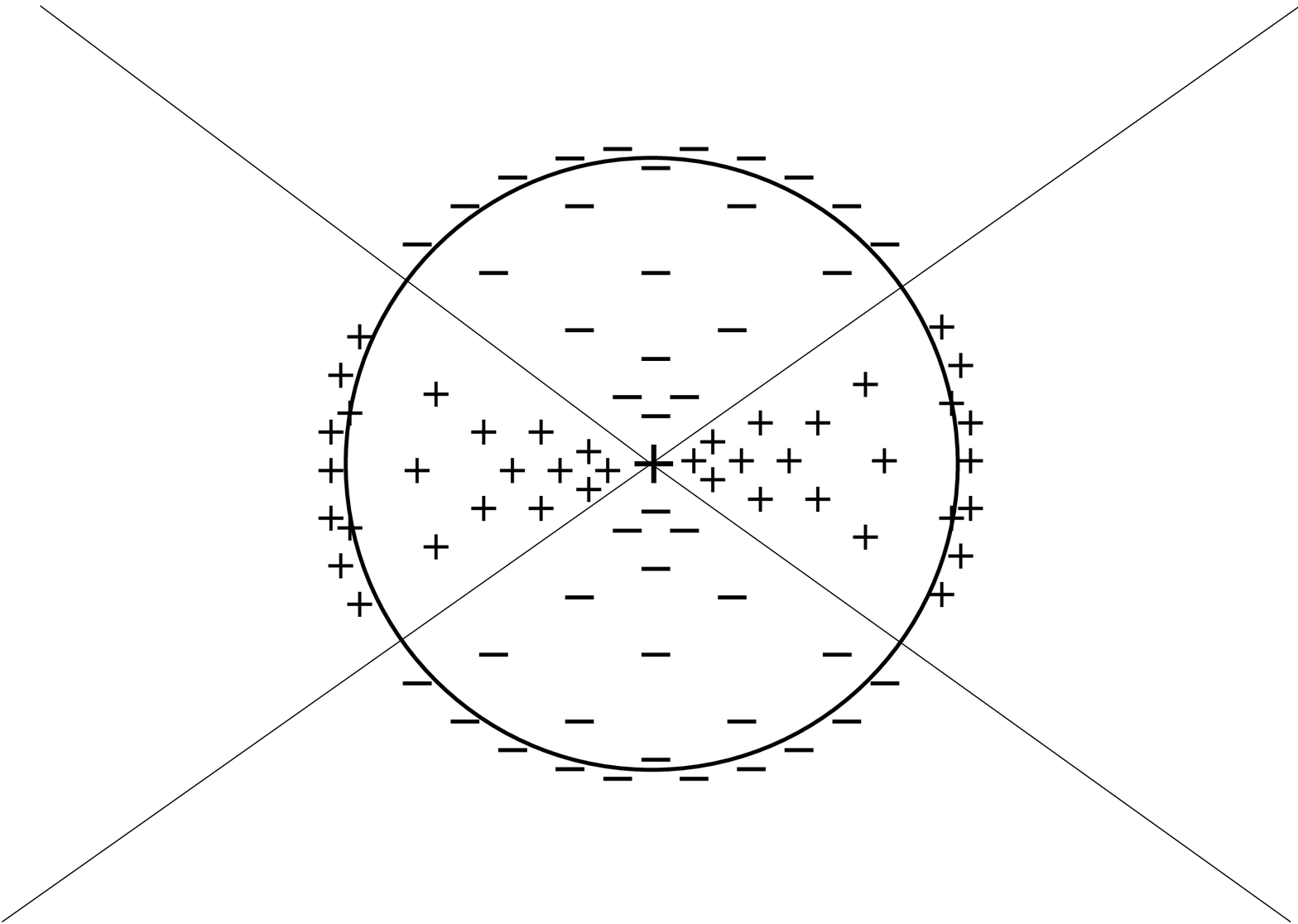}
\caption{\hspace{.1in}Charge distribution 
on an ``aligned rotator'' in the
laboratory.  Here the magnetic dipole is at the center
and there is a central charge (the slightly larger ``+''
at the center), space charge separation in the interior, and
a trapped surface charge.  All three components are required
for ${\bf E \cdot B} = 0$ in the interior.
An arbitrary uniform surface charge could
be added (e.g., to give a total charge of zero, say), since
this charge would contribute no additional internal $\bf E$ field.} 
\label{fig:lab}
\end{minipage}
\hfil\hspace{\fill}
\begin{minipage}{75mm}
\includegraphics[width=75mm]{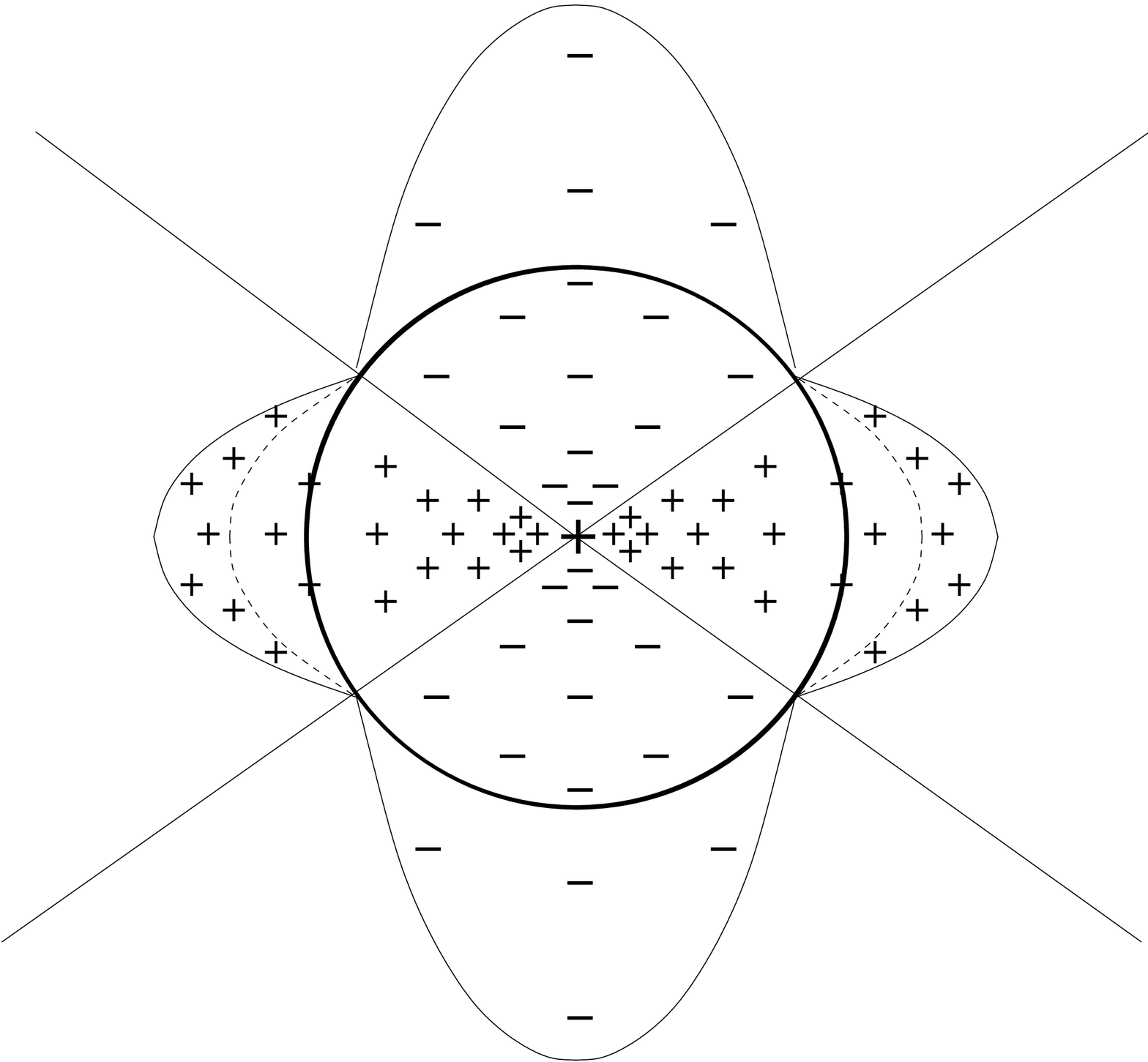}
\caption{\hspace{.1in}Redistribution of
charges in Figure~5 if the charges were to be
released from the surface.  The interior central charge
and interior charge separation is unchanged, while the surface
charge becomes a surrounding electrosphere (here the exact
configuration depends on how much uniform surface charge was
added.}
\label{fig:dometorus}
\end{minipage}
\end{figure}

It is easy to show that you will not get a ``pulsar'' by rotating a
spherical magnet in the laboratory.  It is still true that the system
would accelerate particles off the surface, but the electrostatic
fields (for achievable magnetic fields and rotation rates) obtained
cannot overcome the work function of typical metals and consequently a
surface charge appears but it cannot escape the surface.  This surface
charge is closely connected to the Hones and Bergeson (HB) space
charge.  Inside the star, this space charge is produced (technically
it will be a charge-separation in the conducting material) along with
the central charge.  It is a simple calculation to show that these two
sources of electrostatic field do {\em not} give $\bf {E \cdot B} = 0$
inside the star.  One also needs a third component which is the one
that would have been provided by continuing the space charge on beyond
the stellar surface to ``infinity.''  But there is no such space
charge in the lab.  It could only come from the rotating magnet, but
the work function prevents that.  So what the system has to do is
instead accumulate enough {\em surface} charge to {\em do the same
job.}  What the interior of the star needs is three sources of
electric field:, (1) the central charge, (2) the internal
charge-separation, and (3) a surface charge that provides exactly the
same internal quadrupole electrostatic field that would have been
supplied by filling the surroundings (since this is impossible) with
the HB space charge.  But this field can be provided just as well with
a surface charge of the correct strength and distribution.

This is the point at which the Goldreich Julian model first and fatally
goes wrong.  There is no ``obligation'' that the space charge fill space
around the rotating magnet.  It is just an alternative solution that
must be discarded if the physics will not allow it to happen.  Here the
work function interferes, and the system settles for a surface charge.

But we can always pretend that the work function is zero (in the laboratory,
we could try to blast the surface charges off with laser beams or something).
NOW do we get a pulsar?
No, we get what is shown in Figure~6.

What has happened here is that, as advertised, surface charge released
from the star goes out into the surroundings.  But the surroundings are
tightly constrained.  The particles in the equatorial belt (assumed to
be positive for definiteness) will simply follow magnetic field lines
out, and since they are tightly curved near the star, we can only form
a narrow torus in this way.  The charges above the polar caps could in
principle follow magnetic field lines to ``infinity,'' but they
encounter the trapping region.  Whether or not any charges might make
it to infinity is irrelevant since escaping particles only increase
the total charge of the system, which makes it ever harder for further
particle loss.  All this says is that charge is conserved.  If
positive particles are all trapped in a torus, we cannot very well
continuously lose negative charges.  This criticism of the GJ model
was well publicized (Michel, 1982) and the exact resolution of it
(formation of trapped domes and torus) was suggested in Michel (1980),
demonstrated explicitly in Krause-Polstorff and Michel (1985a,b: KPM),
and summarized in Michel (1991a).  Recent repeat simulations, plus a
number of other numerical experiments (Smith, et al.,
2001: STM) are fully consistent with the original ones.  The generic
distribution of plasma around an aligned rotator is therefore simply a
dome and torus of oppositely charged particles.  For the most part,
this (finite) space charge density is that predicted by Hones and
Bergeson back in 1965.

If one compares what should naturally be found over the magnetic polar
caps (domes of trapped plasma) with what is so generally assumed even
today (regions dominated by outgoing beams of accelerated
particles), we see that the mismatch is shocking.  Again, without a
clear idea of how pulsars actually function, we cannot prove that the
domes are not somehow stripped off and replaced by beams.  But there is
a big difference between wishing for something and actually having it.

\subsection{Expanded work}

It is awkward to insist that one's own work is definitive, so we
should note that the basics of this work have been repeated a number
of times by others independently (Neukirch 1993, Petri et al. 2002,
Shibata 1989, Spitkovsky and Arons 2002, Thielheim and Wolfsteller
1994, Rylov 1989, Zachariades, 1993).  Indeed, this duplication points
up the sociological point made in the Historical Notes: theory has had
a tendency towards competition more than cooperation.  Not one of
these authors exchanged any communication with us saying that they
were re-investigating the issue of trapped plasma around aligned
rotators, or called our attention to their published results.  
Figure~7 shows a recent work using a completely different
numerical routine that faithfully reproduces the KPM results.

\begin{figure}[h]
\begin{minipage}{90mm}
\includegraphics[width=75mm]{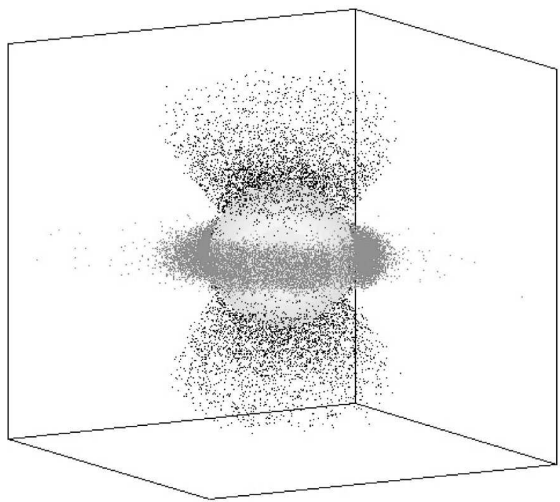}
\end{minipage}
\hfil\hspace{\fill}
\begin{minipage}{75mm}
\caption{\hspace{.1in}Simulations by Spitkovsky
and Arons (2002) reproducing
Krause-Polstorff-Michel results.  Again we obtain two domes
over the dipole=spin axis and a torus of opposite charge
girdling the equator.}
\label{fig:spit}
\end{minipage}
\end{figure}

The original KPM work has recently been entirely redone SMT (2001)
giving complete reproduction of the original results.  This new code
was used to additionally test some other issues.  Two in particular
are of interest, as we will next discuss.

\subsection{GJ model failure}

First, it should be evident from examination of
Figure~6 that not only do the surroundings of an
aligned rotator {\em not} fill with plasma, but that is an impossible
solution if plasma has to follow magnetic field lines.  There is then
no way to get a torus to extend to large distances without having the
particles somehow cross intense magnetic field lines (an issue we will
return briefly to).  At large distances, the positive particles would
have had to come along magnetic field lines from surface areas where
there could only be negative surface charges.  But, we can numerically
put charged particles wherever we want them, so as a numerical
exercise in STM (2001), initial GJ configurations were created out to large
distances and then allowed to relax (since they do not correspond to
natural solutions to the aligned rotator, they are not in static
equilibrium for {\em any} finite size).  When they relaxed, they immediately
fell apart to form domes and tori as shown in
Figure~8 (pre-collapse) and
Figure~9 (fully collapsed).

\begin{figure}
\begin{minipage}{90mm}
\includegraphics[width=75mm]{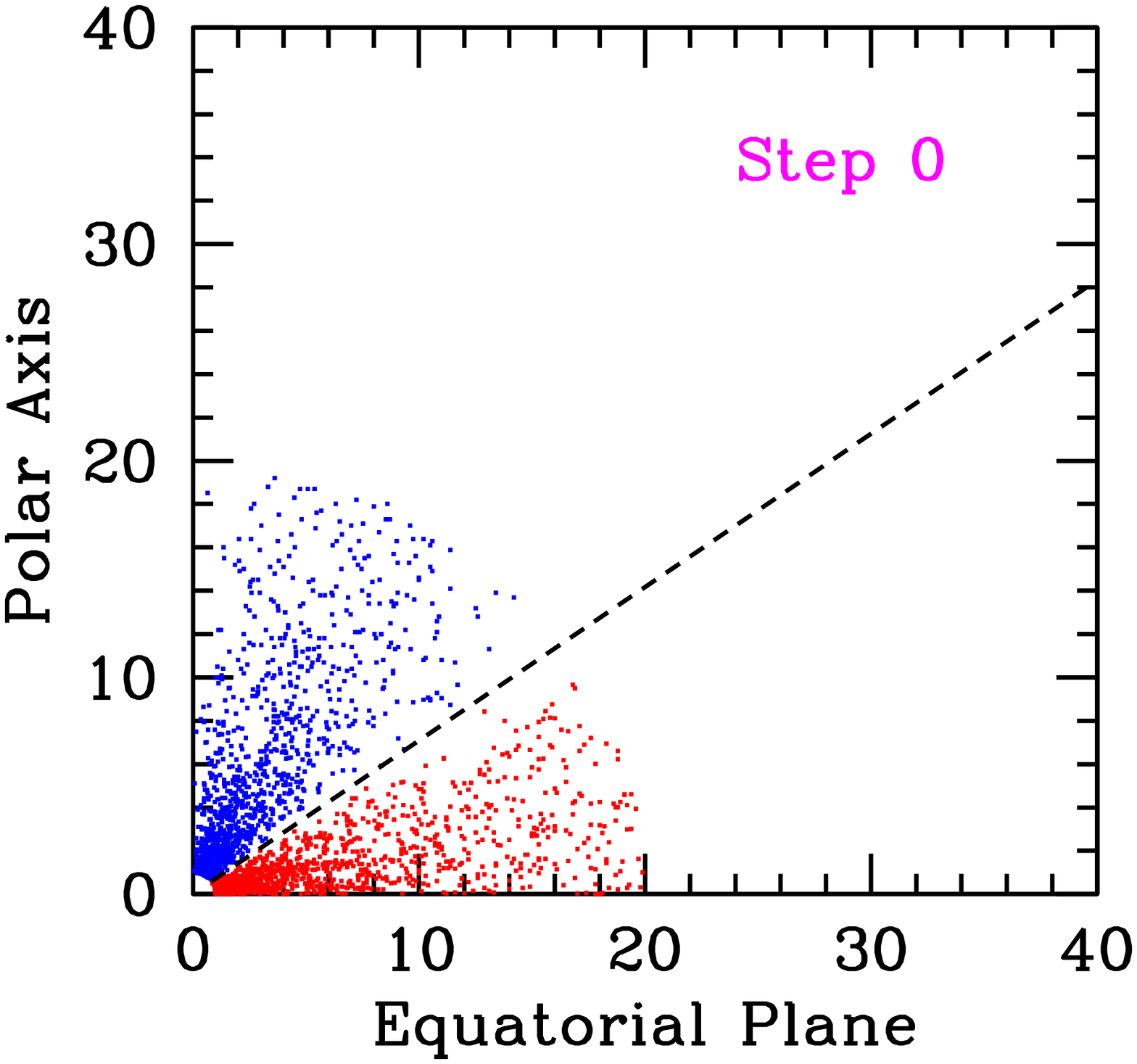}
\caption{\hspace{.1in}Goldreich-Julian configuration
at start of simulation.  The configuration
was arbitrarily truncated at $r=30$, since the problem is largely
scale invariant.  The configuration already looks somewhat like
a dome and torus owing to the HB charge density going to zero along
the dotted line (which therefore separates charges of opposite sign).}
\label{fig:collapse0}
\end{minipage}
\hfil\hspace{\fill}
\begin{minipage}{75mm}
\includegraphics[width=75mm]{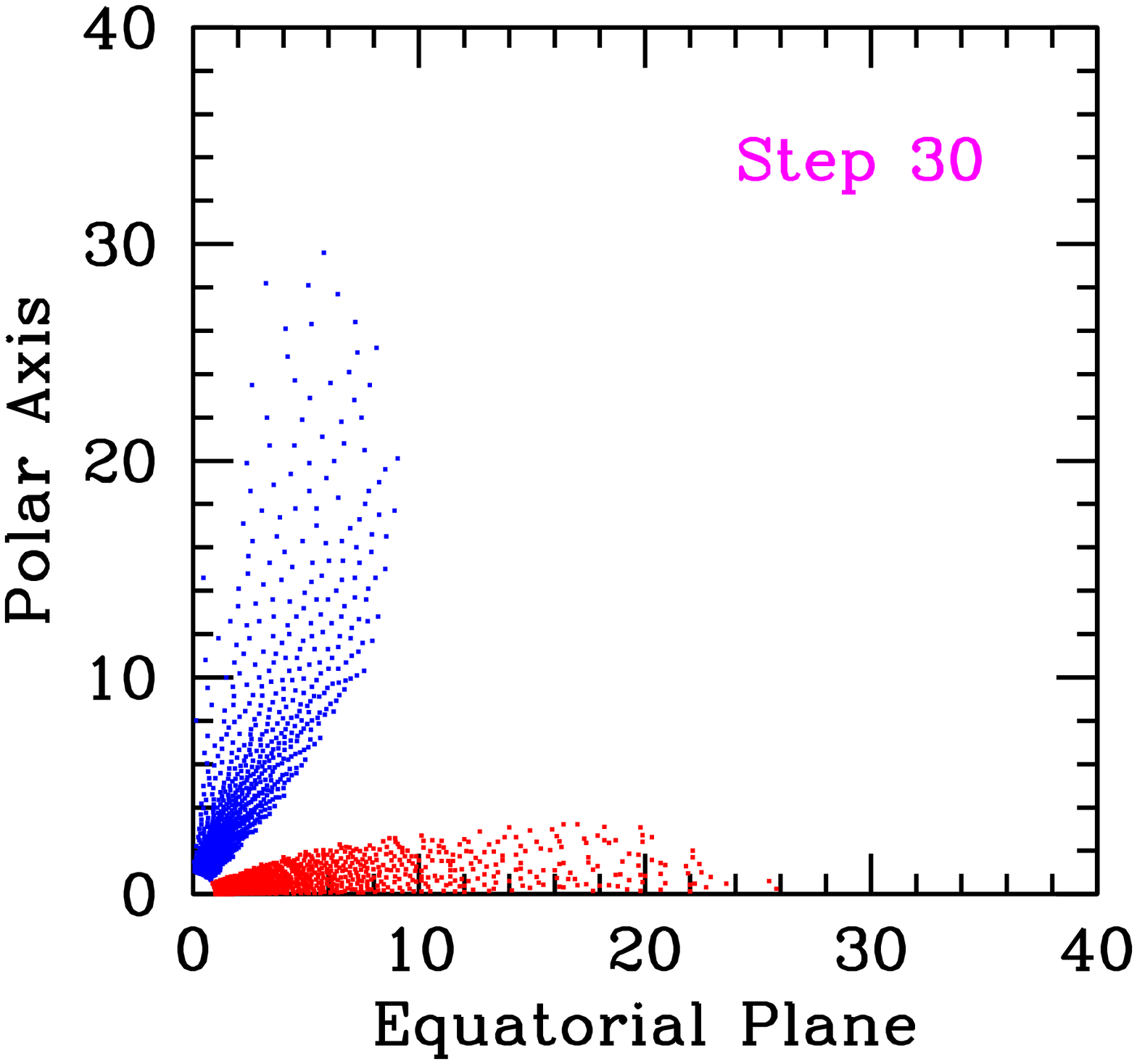}
\caption{Goldreich-Julian configuration
at end of simulation (about 30 steps).
There is no action to pull ``replacement'' charges from the magnetic
polar caps.  A color movie showing the evolution is available at
http://spacsun.rice.edu/~ian/gjpulsar/.  There, blue particles above
and red particles below represent the two signs.}
\label{fig:collapsed30} 
\end{minipage}
\end{figure}

Putting this another way, there is no obligation whatsoever for plasma
to be replaced.  The idea that a ``wind'' would be spun off at the
``light cylinder'' is irrelevant because there is no plasma
extending to this distance (unless of course other physics is put in).
And it is irrelevant even if a wind removed plasma, since there is no
obligation that it be replaced.  {\em The argument that there
must be acceleration of particles from the polar caps to replace loss
at large distances is
baseless}.  If particles are accelerated off polar caps, it will be for
some other reason.  At present, it is a myth.
Historically, it should be noted that even earlier it was shown that
a purely aligned rotator could not work (Scharlemann, et al., 1978),
but this was based on a different physical constraint and did not answer
the question of what would happen if aligned.  
This question was answered in KPM.

\subsection{Pair production}

The possible role of pair production has been an interesting one,
and we find it very attractive (Michel 1991b) because it could explain
how charged bunches would form naturally and thereby account for the
coherency of pulsar radiation.
One role of pair production
might be to provide ionization outside of the neutron star and thereby
help ``fill'' the magnetosphere as imagined in the GJ model (although
these authors were clear in their assumption that the magnetospheric
particles all came from the surface).  Although something like this should
be possible (owing to the huge $\bf {E \cdot B} \neq 0$ regions between
the domes and the tori) STM (2001) show that the consequent filling
of the domes and tori reduce this source and would turn it off.  Moreover,
for typical pulsars where the magnetic field at the famous light cylinder
would be only of the order of a few gauss, pair production would have no
chance of operating.  Pair production was suggested in the first place
only because the pulsar magnetic fields were so large at the surface.

\subsection{FrankenModels}

To reprise, a simple aligned rotator should have domes and a torus.
In between would be a huge vacuum gap.  It was suggested on the basis
of such gaps that high-energy emissions might develop in such gaps
from the Crab and Vela pulsars (Cheng, et al., 1986).  This is exactly
what was just discussed above, some sort of discharge activity between
the dome and torus.  Ironically, when Cheng, Ho, and Ruderman
introduced this gap model for the Crab pulsar, they cited KPM as
justification.

This today has lead to what one might call ``FrankenModels'' after the
pieced-together (from ill-fitting bits) monster in Mary Shelley's
novel.  Here we find models in which radio emission comes as ever from
the polar caps, and {\em additionally} high-energy emission comes from
vacuum gaps.  But the gaps, {\em contradict} the idea that there is
activity just above the polar caps.  And if there were activity over
the polar caps, it would presumably be because the gaps had been
filled in and (somehow) permit GJ activity.  A classic example
of wanting to eat the cake and still have it too.

\subsection{Custom Gaps}

In much the same way, the fact that CHR more or less postulated gaps
has lead to the idea that gaps are somehow a customizable feature that
can be added in whatever size and location that would be handy.  This
of course is completely at odds with the search for models that are
physically self-consistent.  Some of this seems to stem from an early
paper by Holloway (1973) where a simple explanation for the possible
existence of gaps is given (Michel 1991a).  This paper is usually
cited by proponents of gamma-ray emission coming from gaps, which is
physically inconsistent: the reason for such gaps is that they cannot
fill {\em unless} a local source of ionization were present (i.e., a
putative gamma-ray generation activity), which would then close them
and turn off the activity.  See ``FrankenModels,'' above.  Holloway
never seems to have realized, as we (and others now) claim, that the
gaps vitiate the GJ model, but assumed that the gap only consisted of
a tiny separation along the zero charge line (locus of $\bf B \cdot \Omega$)
(see Holloway and Price 1981).

\section{CENTRIFUGAL PREOCCUPATION}
\vspace{2mm}
Another huge distortion of the likely physics has been the idea that
pulsar action is ruled by centrifugal force.  This idea is certainly a
plausible one if one believes that a plasma wind is ``spun off'' from
a pulsar and ``requires'' the replacement of plasma from the magnetic
polar caps.  But as we have seen, neither a centrifugal wind nor an
acceleration region to replace the wind particles results from the
basic physics.  Worse, it is essentially impossible to drive currents
of both signs with centrifugal force.  This conclusion should not be a
surprise.  After all, centrifugal force is an inertial force that acts
on bodies proportional to their mass and independent of their other
properties (e.g., charge).  The problem is that plasma surrounding
even an aligned rotator does not ``know'' that the star is spinning.
All the charged plasma particles experience are the local
electromagnetic forces.  Of these forces, the electrostatic force is
dominated by the monopole moment of the central charge, which drops
off as
\begin{equation}
E_r \approx \frac{1}{r^2}
\end{equation}
while the magnetic field will be dominated by the dipole
\begin{equation}
B_\theta \approx \frac{1}{r^3}
\end{equation}
and consequently the drift velocity will be azimuthal and scale as
\begin{equation}
V_\phi = \frac{E_r}{B_\theta} \approx r .
\end{equation}
The ``light cylinder'' will be whatever the distance $r$ is that
$V_\phi = c$.  

So far, so good, but what about the ``centrifugal forces spinning off
a wind?''  The problem is that once beyond this magical distance, the
electric field will dominate for the simple reason that particles cannot
go faster than $c$.  But a monopole will accelerate away only one sign
of charge.  And charge conservation forbids endlessly spinning off only
one sign of charge.  So the whole notion that centrifugal winds will spin
off a wind collapses.  Also the idea that the centrifugal forces become
``infinite'' at the light cylinder distance is physical nonsense.  The
largest forces will be of the order of $eE$.

\subsection{Wave Fields}

The simplest resolution to the problem (of how charges of both signs
can be lost) is that the dominant forces are really the
electromagnetic wave fields.  It comes as no news to the observational
community that the neutron star magnetic fields should be inclined to
the spin axis.  How else to have ``the searchlight?''  The idea of
centrifugal dominance comes only because it is assumed that the
aligned rotator would function as a pulsar.  If not, inclination would
also be necessary to activate the pulsar action.  Indeed, it is
unlikely that the magnetic field of a neutron star could be so
perfectly aligned and purely dipolar that there would be no wave fields
whatsoever at the wave zone.

The effect of wave fields is to accomplish that which is incorrectly 
attributed to centrifugal forces: drive off a wind of either charge.
In a wave field we have
\begin{equation}
E_\phi \approx \frac{1}{r}
\end{equation}
and 
\begin{equation}
B_\theta \approx \frac{1}{r}
\end{equation}
giving
\begin{equation}
V_r \approx \frac{E_\phi}{B_\theta} \rightarrow c ,
\end{equation}
which is independent of charge.

Unlike centrifugal forces, there is a quasi-spherical {\em wave zone}
at essentially the same distance as the light cylinder, but no cylinder.
From an inclined pulsar, circularly polarized waves race up the
spin axis to accelerate out plasma there as well as out the equatorial
axis.
If we examine the Crab pulsar, as shown in Figure~10,
we see that the overall topology is essentially that of a torus and
two domes, with this now representing a wind, so that the domes now
appear to be jets shooting out of the spin axis.  

\begin{figure}[h]
\begin{minipage}{90mm}
\includegraphics[width=75mm]{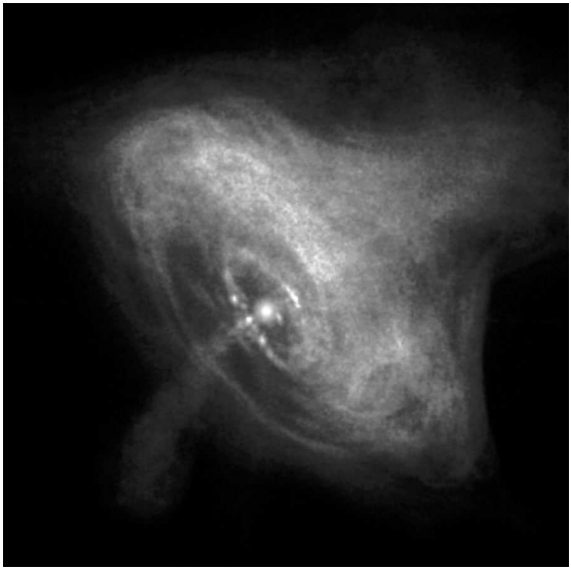}
\end{minipage}
\hfil\hspace{\fill}
\begin{minipage}{75mm}
\caption{\hspace{.1in}X-ray image of the
Crab Nebula from CHANDRA.  Note how
the equatorial distribution of wisps and the polar jets exactly mimic
the domes and torus distribution expected about a rotating neutron star.}
\label{fig:crab}
\end{minipage}
\end{figure}
If correct, this interpretation would suggest huge charge separations
driven by equatorial outflows of one sign of charge and polar jets of
the other sign, which in turn might account for the one jet that is
clearly starting to turn back to the nebula at its outer reaches.

\section{``NEW'' IDEAS}
\vspace{2mm}
As noted in the Historical Notes, 35 years is too long for anyone to
assume that an idea new to them is new to the field.  Some explicit
examples follow.

\subsection{Pulsar Disks}
\begin{figure}[t]
\begin{minipage}{90mm}
\includegraphics[width=75mm]{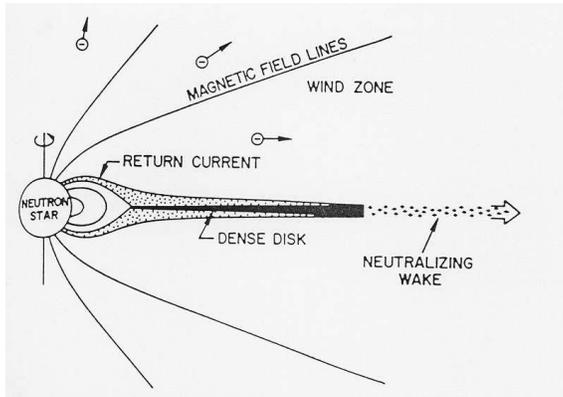}
\end{minipage}
\hfil\hspace{\fill}
\begin{minipage}{75mm}
\caption{A material disk surrounding a rotating neutron star,
which would provide an electrical short through the closed field lines.
Activity would be essentially ensured.}
\label{fig:disk}
\end{minipage}
\end{figure}

Recently, a spate of papers have appeared suggesting that timing
inconsistencies in pulsars could be accounted for by a ``fossil'' disk
around the neutron star (ordinary albeit degenerate matter, as opposed
to a disk of nonneutral plasma), as illustrated in
Figure~\ref{fig:disk} (Alpar et al., 2001, Menou et al., 2001, 
Marsden et al., 2001, Perna et al., 2000).
But not only is this not a new idea (Michel
and Dessler, 1981, 1983, 1985, and Michel, 1991a), but even the
putative source of the disk (supernova fall back) is not even new
(Michel 1988).  A fossil disk is one of the few dynamic elements that
could be added to a neutron star without being excluded by timing
measurements.

This idea actually originated in an attempt to explain pulsar action
in view of the failure of the GJ model.  A disk would be natural
in that one would have two elements in an electric circuit (the
neutron star and the disk) that could not be corotating and therefore
could not be at the same potentials.  Some sort of discharge would
certainly be plausible, and the later papers addressed issues such as
``could the disk last along enough.''  The term ``fossil'' disk is itself
not even new.

Presumably it is only a matter of time before someone is bold enough
to suggest that disks are indeed needed for pulsar action per se.,
which will then not be a new idea.

\section{Concluding Remarks}

A number of new physical issues are coming to the fore.

\subsection{Nonneutral Plasmas}

Most astrophysicists or physicists are not taught nonneutral
plasmas.  There are relatively few sources
Davidson (1990), Michel and Li (1999), Michel (1982, 1991a).
Yet nonneutral plasmas are the natural plasmas to be found around
strongly electrified objects like rotating neutron stars, simply because
the huge electric fields tend to stratify the plasma
(as in producing domes and tori) and selectively pull
out plasma (from conducting surfaces) of only one sign.

The fact that the plasma surrounding a rotating magnetized neutron star
should be arranged in the form of domes and tori should be understood as a
fundamental one which would have to be explicitly modified if one were to find
a structurally different configurations (such as accelerating gaps over
the polar caps) instead of simply being ignored because it doesn't fit
preconception.

\subsection{Importance of Large Amplitude Electromagnetic Waves}

As discussed in Michel and Li (1999), wave acceleration can be hightly
efficient {\em along} magnetic field lines such as along the spin axis
of a pulsar.  The energetic jets see from the Crab pulsar are quite
possibly evidence of this basic physical fact.  Such jets would not be
due to astrophysical ``nozzles,'' as is almost universally the assumption,
but rather do to preferential acceleration where the magnetic fields do
not form obstacles (i.e., along magnetic poles).

\subsection{Danger when theorists start looking at the same page}

Theorists seem uninterested in why 35 years have passed with so little
success.  Not even interested enough when it can be shown that the
favorite model is a cartoon model not based on real physics.  The possible
bad news here is that a number of other people have now become interested
in how nonneutral plasmas impact our understanding of pulsars.  Then
everyone might have to learn this stuff!

\section{ACKNOWLEDGMENTS}
This work was supported in part by NASA under STSI grant HST-G)-7407.04
and under Smithsonian grant G01-2076C.

\end{document}